\documentclass[journal=jpcbfk,manuscript=letter,layout=traditional]{achemso}
\usepackage{amsmath,amsfonts,amssymb}
\usepackage{fullpage}
\usepackage{setspace}
\usepackage{hyperref}
\usepackage{graphicx}
\usepackage{color}
\usepackage{bm}	
\usepackage{upgreek}
\usepackage{comment}

\newcommand{\onlinecite}[1]{\hspace{-1 ex} \nocite{#1}\citenum{#1}} 

\title{Dynamic Monte Carlo simulation of coupled transport through a narrow multiply-occupied pore}

\author{Dezs\H{o} Boda}
\email{boda@almos.vein.hu}
\phone{+36-88/62-4325}
\fax{+36-88/62-4548}
\affiliation[University of Pannonia]{Department of Physical Chemistry, University of Pannonia,  P. O. Box 158, H-8201 Veszpr\'em, Hungary}

\author{\'Eva Cs\'anyi}
\affiliation[University of Pannonia]{Department of Physical Chemistry, University of Pannonia,  P. O. Box 158, H-8201 Veszpr\'em, Hungary}
\alsoaffiliation[Edison House]{\\Edison House Holding Co., Ltd., H-2724 \'Ujlengyel, Hern\'adi u. 18., Hungary}

\author{Dirk Gillespie}
\affiliation[Rush University Medical Center]{\\Department of Molecular Biophysics and Physiology, Rush University Medical Center, Chicago, IL 60612}

\author{Tam\'as Krist\'of}
\affiliation[University of Pannonia]{Department of Physical Chemistry, University of Pannonia,  P. O. Box 158, H-8201 Veszpr\'em, Hungary}

\date{\today}

\begin{document}

 \onecolumn
\newpage
\begin{abstract}
\small
Dynamic Monte Carlo simulations are used to study coupled transport (co-transport) through sub-nanometer-diameter pores. 
In this classic Hodgkin-Keynes mechanism, an ion species uses the large flux of an abundant ion species to move against its concentration gradient. 
The efficiency of co-transport is examined for various pore parameters so that synthetic nanopores can be engineered to maximize this effect. 
In general, the pore must be narrow enough that ions cannot pass each other and the charge of the pore large enough to attract many ions so that they exchange momentum. 
Co-transport efficiency increases as pore length increases, but even very short pores exhibit co-transport, in contradiction to the usual perception that long pores are necessary. 
The parameter ranges where co-transport occurs is consistent with current and near-future synthetic nanopore geometry parameters, suggesting that co-transport of ions may be a new application of nanopores.
\end{abstract}

\begin{center}
\textbf{TOC graphic:}\\
\vspace{0.5cm}
\rotatebox{0}{\scalebox{0.6}{\includegraphics*{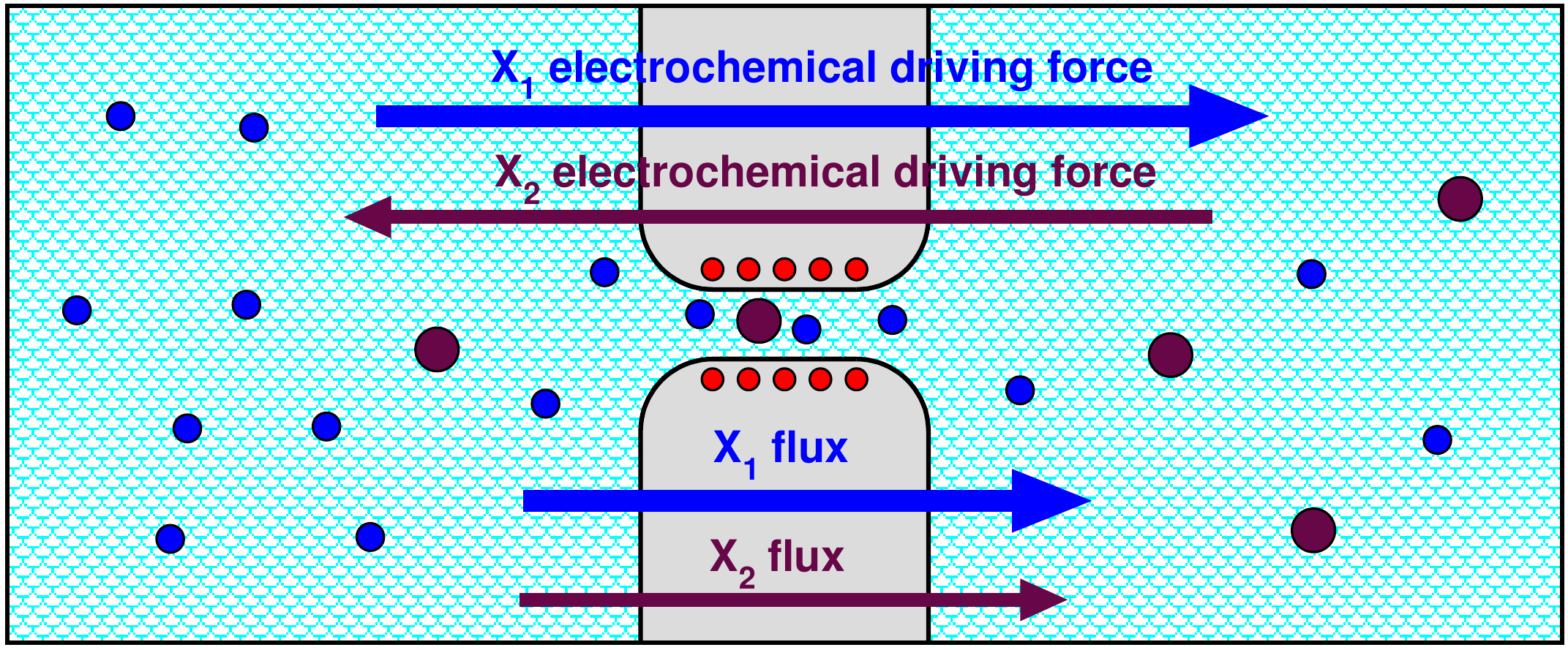}}}
\end{center}

\begin{center}
\end{center}

\maketitle

%%%%%%%%%%%%%%%%%%%%%%%%%%%%%%%%%%%%%%%%%%%%%%%%%%%%%%%%%%%%%%%%%%%%%
%%%%%%%%%%%%%%%%%%%%%%%%%%%%%%%%%%%%%%%%%%%%%%%%%%%%%%%%%%%%%%%%%%%%%

\newpage

\section{Introduction}
\label{sec:intro}

Biological studies have shown that it is possible to move molecules ``uphill'' against their electrochemical potential gradient by coupling their flow to the large downhill flow of another particle species.\cite{hodgkin_jp_1955,defelice_arp_2007} 
The coupling of fluxes occurs when the momentum of the ions is coupled as they flow through narrow multiply-occupied sub-nanometer-sized pores. 
In the biological cases, these pores can be ion channels or channel-like transporters.\cite{Hille,defelice_arp_2007} 
Recently, however, synthetic nanometer-diameter pores have been engineered in a variety of materials including PET, silicon nitride, and polycarbonate \cite{howorka_csr_2009}.
Moreover, ion channels have been inserted in these pores to make them even more narrow \cite{hall_nnt_2010,kocer_bsbe_2012}, even to the point of the single-file motion of ions required for momentum coupling\cite{balme_nl_2011}. 

The purpose of this paper is to simulate this kind of coupled transport (co-transport) directly with a molecular simulation method for the first time over a wide range of experimental and fabrication parameters. 
Specifically, our goal is to understand the general mechanisms behind co-transport by studying the effect of various factors that influence coupling of movements of two ionic species in a narrow multiply-occupied pore. 
With a general understanding of what parameters enhance uphill ion transport it will be possible to fabricate synthetic nanopores and nanoporous materials for a wide range of applications like low-level contaminant removal, analyte concentration amplification, and energy storage. \cite{rogers_book,baker_2012}

Ions must be moved against their electrochemical gradient using energy. 
In biological cells, this external energy can be a direct chemical energy (ATP hydrolysis in the case of ATPases, for example) used for conformational changes in the transporter proteins that transport the ions bound on one side to the other side during this structural change.
Pumps working this way maintain the concentration gradients of various ions (Na$^{+}$, H$^{+}$, K$^{+}$) that can be harvested in co-transport as a secondary energy source. 
Other mechanisms of co-transport have also been proposed. 
For example, Eisenberg and coworkers\cite{chen_bj_1993,eisenberg-1996-1} suggested the possibility that ion fluxes can be coupled through the electric field.

Here, we investigate the narrow-pore mechanism where one uses an existing concentration gradient (e.g., of Na$^{+}$) to create a large flux to push another species (that is usually present at much smaller concentration) with the flow against its own concentration gradient. 
Calculations based on various kinetic models have been performed on the basis of this model \cite{defelice_bs_2001}, but direct computer simulations using a molecular model that explore the co-transport mechanism in detail, to our best knowledge, are still absent. 
Several other groups, including those of Coalson \cite{graf_jpcb_2000,graf_jpcb_2004,cheng_jpcb_2005}, Chung \cite{hoyles-cpc-115-45-1998,chung-bj-75-793-1998,chung-bj-77-2517-1999,corry-bj-2000,corry-bj-2001,corry-bj-82-1975-2002,chung-bba-1565-267-2002,corry-bj-84-3594-2003,Corry_BBA_1}, and Roux \cite{im_bj_2000,noskov-bj-87-2299-2004,allen_jgp_2004,luo_jpcb_2010,egwolf_jpcb_2010,lee_jcc_2011} used various simulation methods (including Brownian dynamics, dynamic lattice Monte Carlo, and molecular dynamics) to study diffusion of ions through narrow biological ion channels. 
Although the simulation techniques used by these groups could handle microscopic coupling between ions (interactions through intermolecular potentials or collisions), none of these groups studied the macroscopic phenomenon of coupled transport systematically, as is done in this paper.

On the other hand, Chou and Lohse \cite{chou_jcp_1998,chou_prl_1999} did consider co-transport in single-file pores. Using a simple one-dimensional lattice exclusion model, they found single-file coupled transport through model zeolites and channels using kinetic models in a lattice simulation. However, given the simplicity of the model, the need for molecular dynamics or Monte Carlo (MC) simulations to reveal microscopic details was raised by the authors.

In this work, we build a simple molecular model for a co-transporting pore, in which a narrow pore is lined with a number of structural charges to attract a sufficient number of cations into the pore.
This makes the pore multiply-occupied. 
This, and the narrowness of the pore, makes the movement of ions coupled because the ions cannot pass each other.

%%%%%%%%%%%%%%%%%%%%%%%%%%%%%%%%%%%%%%%%%%%%%%%%%%%%%%%%%%%%%%%%%%%%%%%%%%%%%%%%%%%%%%%%%%%%%%%%%%%%%%%%%%%%%%%%%%%%%%%%%%%%%%%%%%%%%%%%%%%%%%%%%%%%%%%%%%%%%%%%%%%%%%%%%%%%%%%%%%%%%%%%%%%%%%%%
%%%%%%%%%%%%%%%%%%%%%%%%%%%%%%%%%%%%%%%%%%%%%%%%%%%%%%%%%%%%%%%%%%%%%%%%%%%%%%%%%%%%%%%%%%%%%%%%%%%%%%%%%%%%%%%%%%%%%%%%%%%%%%%%%%%%%%%%%%%%%%%%%%%%%%%%%%%%%%%%%%%%%%%%%%%%%%%%%%%%%%%%%%%%%%%%

\section{Model and Method}
\label{sec:model_method}

\begin{figure*}[t]
\begin{center}
\rotatebox{0}{\scalebox{0.6}{\includegraphics*{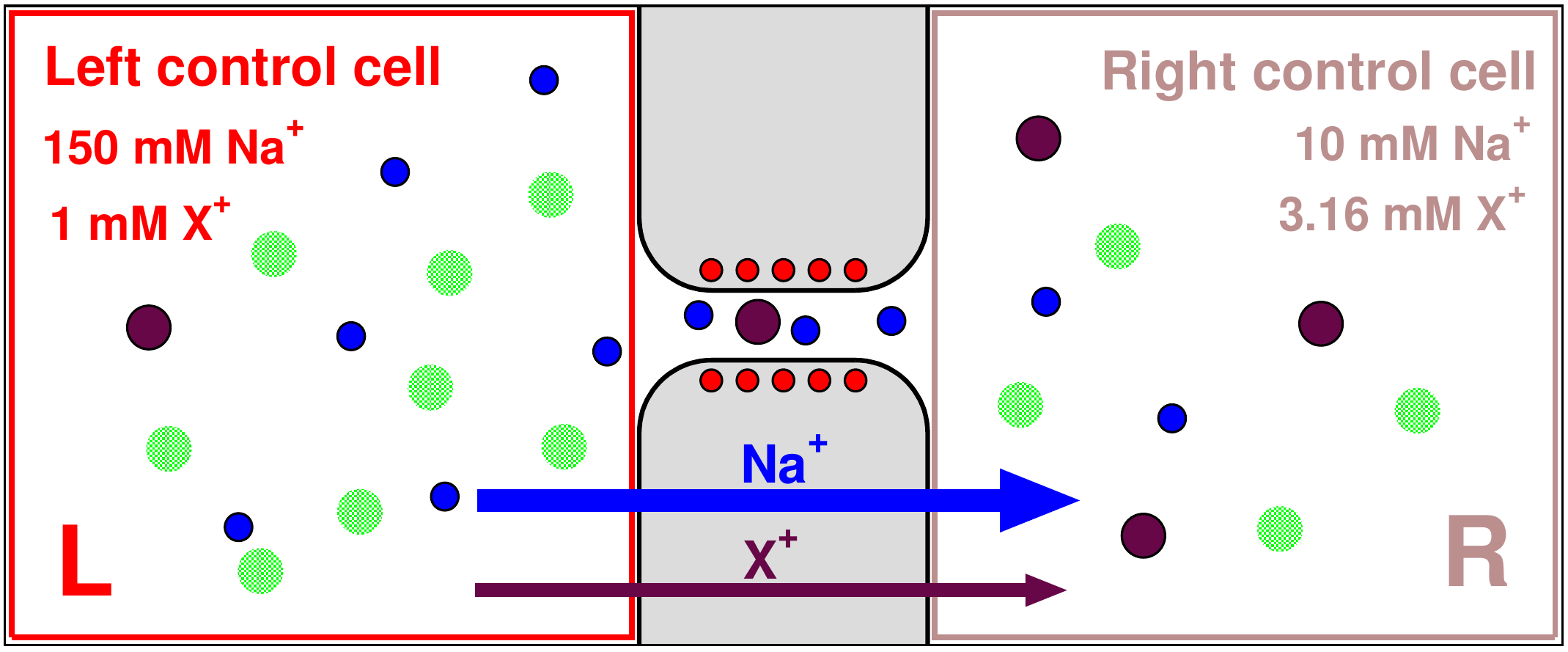}}}
\end{center}
\caption{\small Geometry of the simulation cell. 
A membrane with a pore is placed in the center between two control cells. 
The central cylindrical part of length $H$ and radius $R_{\mathrm{p}}$ is surrounded by partial charges placed in rings (red circles).
The sum of the partial charges is $Q$.
The figure shows ionic concentrations in the control cells for a typical case, when there is a large concentration difference (150 mM to 10 mM  L$\rightarrow$R) for Na$^{+}$ (blue circles) resulting in a large Na$^{+}$-flux from L to R (M denotes mol/dm$^{3}$ throughout this paper).
The flux of X$^{+}$ ions (brown circles) is coupled to the flux of Na$^{+}$ (indicated by the arrows) against its concentration gradient (3.16 mM to 1 mM R$\rightarrow$L). 
Green circles represent Cl$^{-}$ ions.
\label{fig1}}
\end{figure*}

Our pore model has rotational symmetry obtained by rotating the shape of \ref{fig1} around the centerline.
The pore has rounded edges at the entrances of 0.5 nm curvature radius forming vestibules to the central cylindrical region (of length $H$ and radius $R_{\mathrm{p}}$) of the pore.
This central region is surrounded by $n=(H/0.25\,\mbox{nm})+1$ rings of negative point charges (red circles in \ref{fig1}).
There are four charges in each ring, each in $R_{\mathrm{p}}+0.14$\,nm radial distance from the centerline.
The magnitude of each partial charge is $Q/(4n)$ so their sum gives a predefined value, $Q$.
The total thickness of the membrane is $H+1$ nm.

We use Na$^{+}$ for the abundant species in this study.
The ions are modeled as charged hard spheres, with diameters 0.19 and 0.362 nm for Na$^{+}$  and Cl$^{-}$, respectively.
The species to be co-transported (denoted by X$^{+}$) has a diameter 0.3 nm.
Water is modeled as a dielectric continuum with dielectric constant $\epsilon =78.5$ throughout the system.

The two regions outside the membrane on both sides represent the two baths denoted left (L) and right (R).
In a biological situation, these compartments may be the extracellular and intracellular spaces.
In a technological situation, they are rather called feed and permeate sides. 
The electrochemical driving force for the passive diffusion of the ions is the difference of concentrations in the two control cells. 
Since we enforce electroneutrality in these cells on average, the mean electrical potential across the membrane is 0 V (see Appendix \ref{sec:appendix}).
This is an approximation of a large, well-stirred bath. 
Equivalently, it is as if an electrode were keeping the membrane potential at 0 V, except that our baths remove the excess charges instead of the electrodes.

The transport of ions is simulated by the Dynamic Monte Carlo (DMC) method \cite{rutkai_jcp_2010}, where ions are randomly displaced within a maximum displacement.
The move is accepted or rejected according to the usual MC acceptance criterion (50-150 billion of such moves were attempted).
In the crowded pore, the dominant mechanism of coupling is momentum exchange between clashing particles governed by short-range repulsion that is handled in DMC simulations by rejecting configurations where two hard spheres overlap.
We have also performed molecular dynamics simulations in some cases to verify that DMC correctly captures this phenomena (data not shown).
It has been shown that DMC is an appropriate method to compute relative fluxes in mixtures\cite{rutkai_jcp_2010}, as demonstrated in previous DMC simulations for ion channels \cite{rutkai-jpcl-1-2179-2010,csanyi-bba-1818-592-2012}.
Therefore, we will plot the $J_{\mathrm{X}^{+}}/J_{\mathrm{Na}^{+}}$ flux ratios throughout this paper, which makes sense intuitively because $J_{\mathrm{X}^{+}}/J_{\mathrm{Na}^{+}}$ characterizes the efficiency of co-transport.  
More details can be found in previous papers \cite{rutkai_jcp_2010,rutkai-jpcl-1-2179-2010,csanyi-bba-1818-592-2012} and in the Appendix \ref{sec:appendix}.   

%%%%%%%%%%%%%%%%%%%%%%%%%%%%%%%%%%%%%%%%%%%%%%%%%%%%%%%%%%%%%%%%%%%%%%%%%%%%%%%%%%%%%%%%%%%%%%%%%%%%%%%%%%%%%%%%%%%%%%%%%%%%%%%%%%%%%%%%%%%%%%%%%%%%%%%%%%%%%%%%%%%%%%%%%%%%%%%%%%%%%%%%%%%%%%%%
%%%%%%%%%%%%%%%%%%%%%%%%%%%%%%%%%%%%%%%%%%%%%%%%%%%%%%%%%%%%%%%%%%%%%%%%%%%%%%%%%%%%%%%%%%%%%%%%%%%%%%%%%%%%%%%%%%%%%%%%%%%%%%%%%%%%%%%%%%%%%%%%%%%%%%%%%%%%%%%%%%%%%%%%%%%%%%%%%%%%%%%%%%%%%%%%
%%%%%%%%%%%%%%%%%%%%%%%%%%%%%%%%%%%%%%%%%%%%%%%%%%%%%%%%%%%%%%%%%%%%%%%%%%%%%%%%%%%%%%%%%%%%%%%%%%%%%%%%%%%%%%%%%%%%%%%%%%%%%%%%%%%%%%%%%%%%%%%%%%%%%%%%%%%%%%%%%%%%%%%%%%%%%%%%%%%%%%%%%%%%%%%%
%%%%%%%%%%%%%%%%%%%%%%%%%%%%%%%%%%%%%%%%%%%%%%%%%%%%%%%%%%%%%%%%%%%%%%%%%%%%%%%%%%%%%%%%%%%%%%%%%%%%%%%%%%%%%%%%%%%%%%%%%%%%%%%%%%%%%%%%%%%%%%%%%%%%%%%%%%%%%%%%%%%%%%%%%%%%%%%%%%%%%%%%%%%%%%%%

\section{Results and Discussion}
\label{sec:results}

We first performed a detailed analysis regarding the length, the radius, and the charge of the pore to assess what values are necessary to produce coupled transport of X$^{+}$ and Na$^{+}$.
We started by fixing the length of the pore at $H=1$ nm and changing the charge of the pore at two different pore radii $R_{\mathrm{p}}=0.24$ and 0.48 nm.
There is a large concentration gradient for Na$^{+}$ from L to R (from 150 mM to 10 mM). 
At the same time, there are smaller concentrations of X$^{+}$ in the system (1 mM and 3.16 mM in L and R sides, respectively).
This means that the X$^{+}$ concentration gradient is in the opposite direction of that for Na$^{+}$.

\begin{figure*}[t]
\begin{center}
(A)\rotatebox{0}{\scalebox{0.66}{\includegraphics*{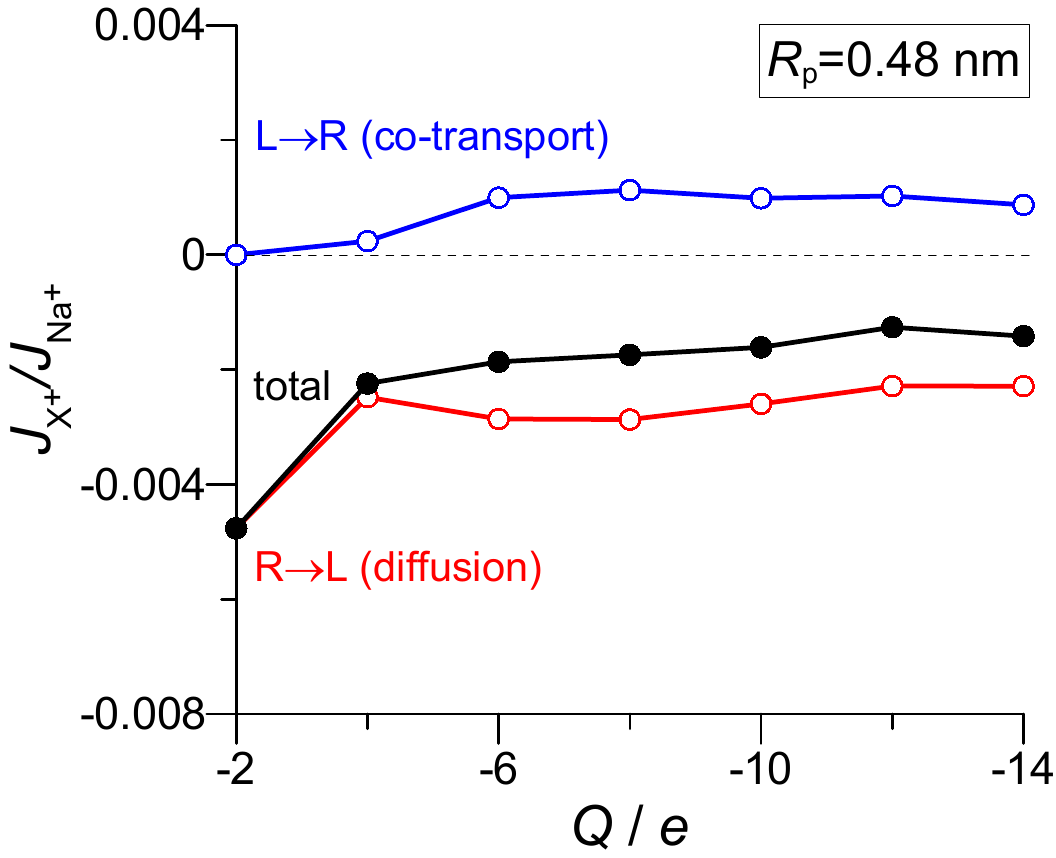}}}\hspace{0.5cm}
(B)\rotatebox{0}{\scalebox{0.66}{\includegraphics*{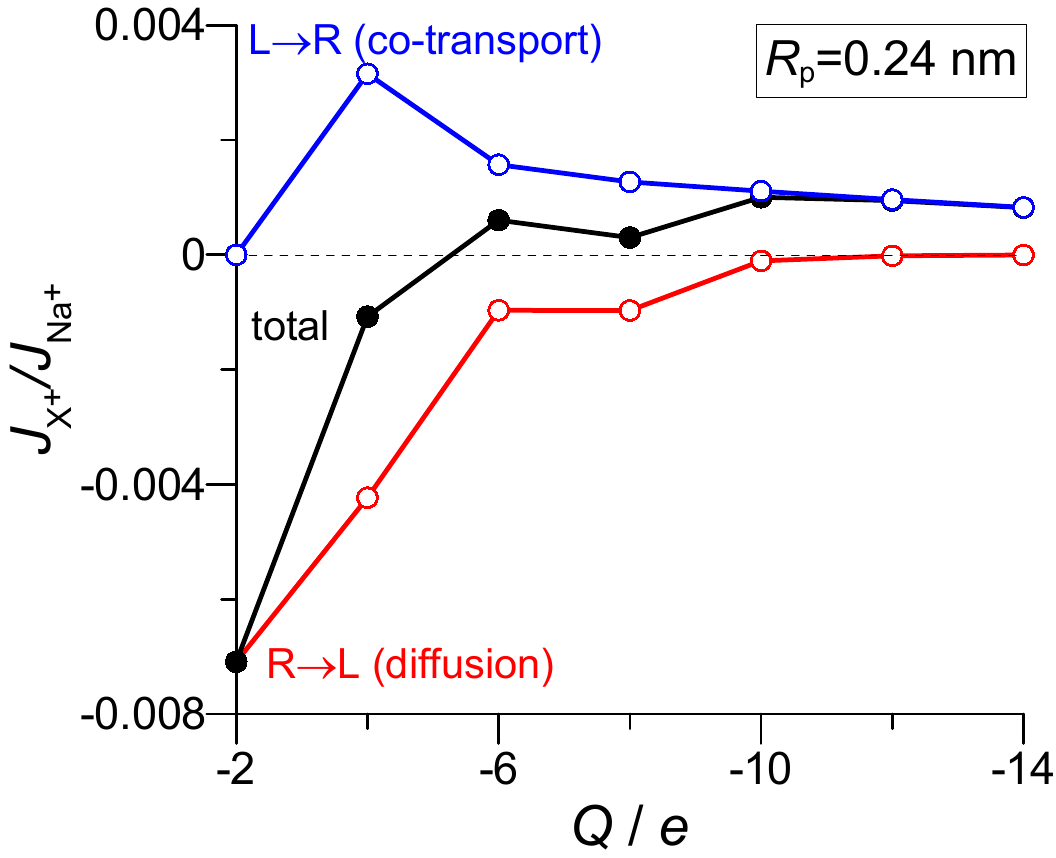}}}
\end{center}
\vfill
\caption{\label{fig2} \small X$^{+}$/Na$^{+}$ flux ratio vs. pore charge, $Q$, for two pore radii $R_{\mathrm{p}}=0.48$ nm (A) and 0.24 nm (B).
Coupled transport occurs when the net flux ratio is positive.
The panels also show the L$\rightarrow$R (co-transport) and R$\rightarrow$L (diffusion) components of the net (or total) X$^{+}$ flux.
The X$^{+}$ concentrations are $[\mbox{X}^{+}]^{\mathrm{L}}=1$ mM and $[\mbox{X}^{+}]^{\mathrm{R}}=3.16$ mM; the ratio is 3.16 (R/L).
The Na$^{+}$ concentrations are $[\mathrm{Na}^{+}]^{\mathrm{L}}=150$ mM and $[\mathrm{Na}^{+}]^{\mathrm{R}}=10$ mM; ratio is 15 (L/R). 
The length of the cylindrical pore is $H=1$ nm, so there are 5 rings of negative charges carrying $Q$ total charge.
}
\end{figure*}

\begin{table}[t]
\renewcommand{\arraystretch}{1}
\renewcommand{\tabcolsep}{2.0mm}
% \begin{center}
\begin{tabular}{c|rrrr}
\hline
\hline
   & \multicolumn{4}{c}{Channel charge / $e$} \\ 
   &	-2   	&	 -6  	&     -10 	&      -14     \\ \hline
00 &	97.66	&	43.11	&	0.66	&	0 \\
01 &	 2.33	&	52.07	&	20.80	&	0.28 \\
02 &	0	&	4.38	&	57.71	&	9.70 \\
03 &	0	&	0.02	&	18.52	&	45.84 \\
04 &	0	&	0	&	0.65	&	37.69 \\
05 &	0	&	0	&	0	&	5.05 \\ \hline
10 &	0.01	&	0.34	&	0.19	&	0 \\
11 &	0	&	0.08	&	1.00	&	0.10 \\
12 &	0	&	0	&	0.43	&	0.59 \\
13 &	0	&	0	&	0.02	&	0.58 \\
\hline
\hline
\end{tabular} 
\caption{\label{tab1} \rm \small The rates (in percent) of various occupancy combinations of X$^{+}$ and Na$^{+}$ in the cylindrical pore for different channel charges, $Q$, for filter radius $R_{\mathrm{p}}=0.24$ nm. 
Other parameters are the same as in \ref{fig2}.
Ion combination $mn$ (1st column) means that there are $m$ X$^{+}$ and $n$ Na$^{+}$ in the cylindrical pore. 
Only those rows are shown, where the rate is larger than 0.1\%.}
% \end{center}
\end{table}

The net flux is a sum of fluxes flowing from R to L (R$\rightarrow$L) and from L to R (L$\rightarrow$R) \cite{}.
(These fluxes correspond to unidirectional fluxes \cite{eisenberg_jcp_1995,eisenberg_cpl_2011} that have been reasonably measured in transport physiology experiments using radioactive tracers. \cite{hodgkin_jp_1955,jacquez_book,rakowski_bj_1989})
The R$\rightarrow$L component is identified with the usual diffusive transport driven by the concentration gradient of X$^{+}$.
The L$\rightarrow$R component is identified with co-transport driven by coupling to the large flux of Na$^{+}$ ions from L to R.
The net effect, therefore, is a result of two competing effects, which is important for our understanding the phenomenon.
(It should be noted that these identifications are artificial.
Even when only diffusion is present without coupling, there are fluxes in both directions.
These concepts, however, promote understanding and serve discussion by seeing where co-transport dominates.)
In \ref{fig2} (and in later figures), therefore, we plot the L$\rightarrow$R and R$\rightarrow$L components in addition to their sum.
Co-transport occurs when the two ionic species go into the same direction.
In this case, the  L$\rightarrow$R component is larger (in absolute value) than the R$\rightarrow$L component, co-transport dominates over diffusion, and the sign of the net flux ratio is positive.

No net co-transport was observed for the wider pore ($R_{\mathrm{p}}=0.48$ nm, \ref{fig2}A). 
The L$\rightarrow$R component is small for every $Q$.
In this case, there is enough space for the ions to travel past each other in opposite directions.
For the narrow pore ($R_{\mathrm{p}}=0.24$ nm, \ref{fig2}B), on the other hand, we observed co-transport when the charge of the pore is large enough.
When there is not enough charge around the pore ($Q=-2e$), there is not enough attraction to attract cations into the pore.
In this case, the pore is not multiply occupied and the movement of Na$^{+}$ and X$^{+}$ cannot be coupled through momentum exchange.
The L$\rightarrow$R component is zero for $Q=-2e$.
Increasing the pore charge, coupling appears, the L$\rightarrow$R component becomes non-zero, and the R$\rightarrow$L component vanishes.
The maximum in the L$\rightarrow$R component at $Q=-4e$ appears because the Na$^{+}$ flux is small in this case, so we normalize with a smaller number.
Increasing $|Q|$ further, Na$^{+}$ flux increases, so the $J_{\mathrm{X}^{+}}/J_{\mathrm{Na}^{+}}$ flux ratio decreases.

The fact that co-transport requires multiply-occupied pore is supported by an analysis where we computed the probabilities of finding various combinations of ions in the pore (see \ref{tab1}).
For $Q=-2e$ pore charge, for example, the pore is empty in 97.66 \% of the time, it contains one Na$^{+}$ in 2.33 \% of the time, and it contains one X$^{+}$ in 0.01 \% of the time.
Obviously, no coupling is possible in this case.
For $Q=-10e$ pore charge, on the other hand, the pore contains only Na$^{+}$ ions most of the time (97.68 \%), but in the remaining time it contains one X$^{+}$ ion next to one or more Na$^{+}$ ions.
In this case, coupling is possible and, together with the confiningly small radius of the pore, it results in X$^{+}$ flux in the opposite direction of its concentration gradient.

\begin{figure}[t]
\begin{center}
\rotatebox{0}{\scalebox{0.66}{\includegraphics*{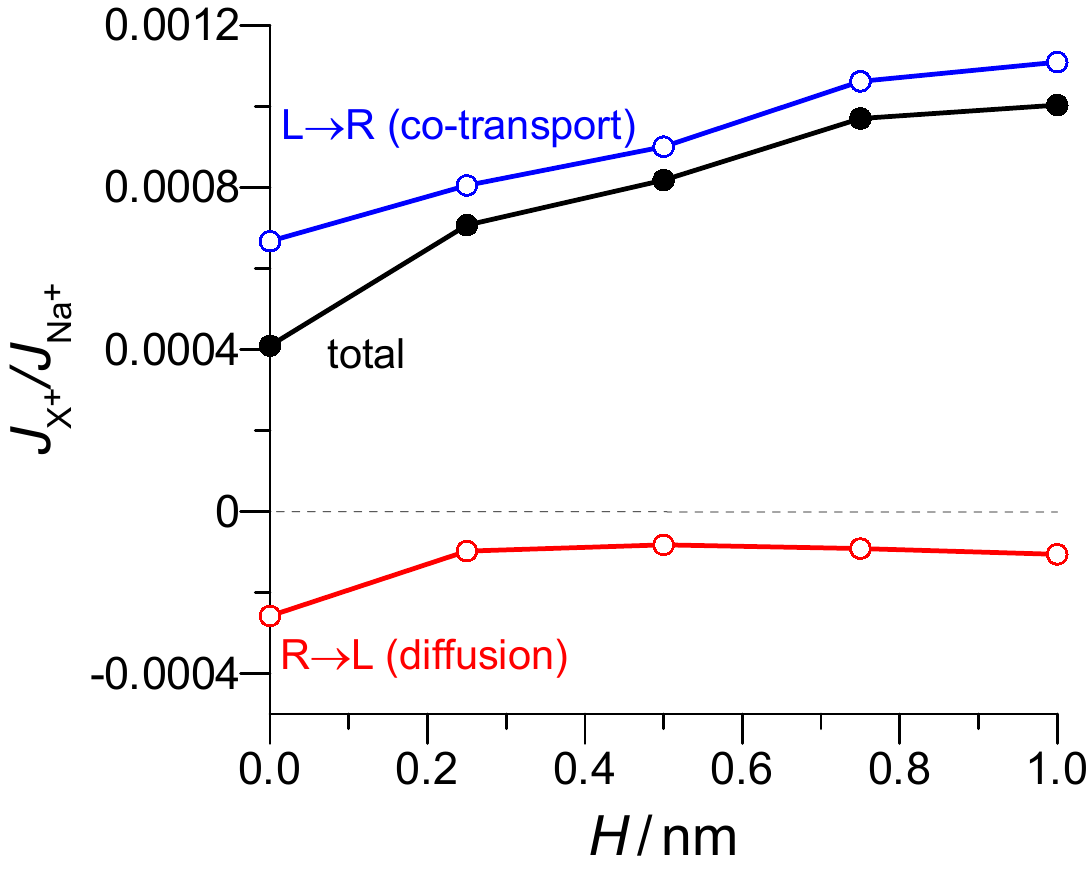}}}
\end{center}
\vfill
\caption{\label{fig3} \small X$^{+}$/Na$^{+}$ flux ratio vs. the length of the cylindrical part of the pore $H$ for pore charge $Q=-10e$ and radius $R_{\mathrm{p}}=0.24$ nm.
The X$^{+}$ and Na$^{+}$ concentrations are the same as in 
% Fig.\ 
\ref{fig2}.
}
\end{figure}

Another interesting aspect is that the flux ratio is saturated; it does not increase further when the pore charge is increased (in absolute value) to extreme values.
Once coupling is established, the flux ratio cannot be increased further by increasing $|Q|$.

\begin{figure}[t]
\begin{center}
\rotatebox{0}{\scalebox{0.75}{\includegraphics*{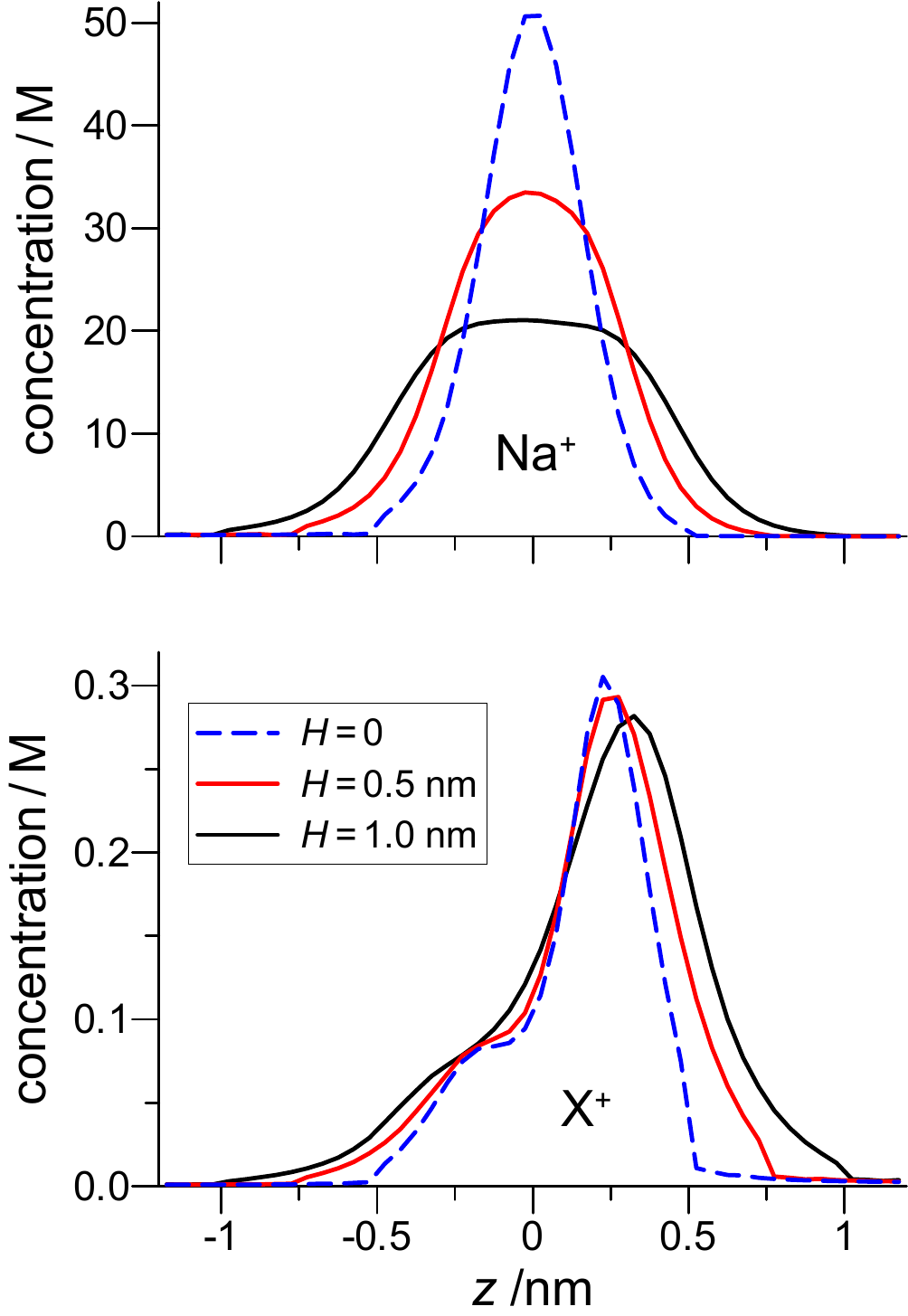}}}
\end{center}
\vfill
\caption{\small \label{fig4}X$^{+}$ and Na$^{+}$ concentration profiles for pore lengths $H=0$, 0.5, and 1 nm, for pore charge $Q=-10e$ and radius $R=0.24$ nm.
The X$^{+}$ concentrations are $[\mbox{X}^{+}]^{\mathrm{L}}=1$ mM and $[\mbox{X}^{+}]^{\mathrm{R}}=3.16$ mM; the ratio is 3.16 (R/L).
The Na$^{+}$ concentrations are $[\mathrm{Na}^{+}]^{\mathrm{L}}=150$ mM and $[\mathrm{Na}^{+}]^{\mathrm{R}}=10$ mM; ratio is 15 (L/R). 
}
\end{figure}

In the next step, we investigated how is this coupling is established.
As seen above, the narrowness of the pore is necessary.
It is believed, because of an experiment by Hodgkin and Keynes \cite{hodgkin_jp_1955}, that coupled transport requires the transport in a long, narrow pore where single filing of ions is forced.
But, how long does the pore have to be to produce coupling?

% Figure 
\ref{fig3} shows the X$^{+}$/Na$^{+}$ flux ratio (and its components) as a function of the length of the narrow cylindrical part of the pore for $Q=-10e$ and $R_{\mathrm{p}}=0.24$ nm.
It is seen that co-transport occurs even if the narrow cylindrical part is absent ($H=0$ nm).
The observed flux ratio is smaller, but the phenomenon is definitely present.
For small $H$, there is more ``leakage'' of X$^{+}$ ions in the ``wrong'' (R$\rightarrow$L) direction, but the L$\rightarrow$R component is sufficient to more than balance it.

Examination of concentration profiles (\ref{fig4}) show that the pore region is crowded even if the cylindrical part is absent ($H=0$ nm); this is ensured by the pore charge ($Q=-10e$).
The important thing is that the ions must crowd in a bottleneck of the pore so that the ion present in abundance (Na$^{+}$) can obstruct the diffusion of the other ion (X$^{+}$) normally driven by its own concentration gradient; that is, Na$^{+}$ ions stand as an obstacle to the movement of X$^{+}$ from R to L no matter whether the pore is long or not.
We find, therefore, that single-filing in a long narrow pore is not necessary to establish coupling between the ions taking part in co-transport.
This result should be taken into consideration when protein structures are analyzed from the point of view of co-transport. 
One does not need a long classical channel; a short but narrow opening suffices to produce coupled transport.

\begin{figure*}[t]
\begin{center}
(A)\rotatebox{0}{\scalebox{0.66}{\includegraphics*{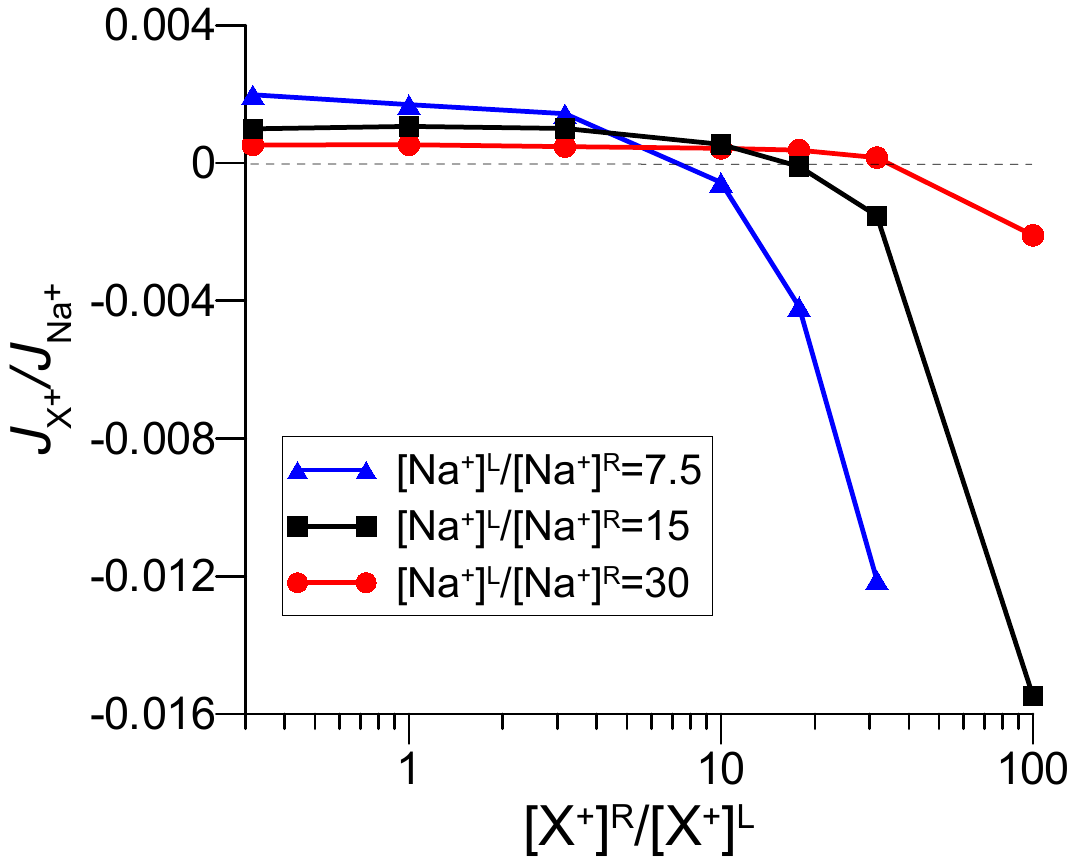}}}\hspace{0.5cm}
(B)\rotatebox{0}{\scalebox{0.66}{\includegraphics*{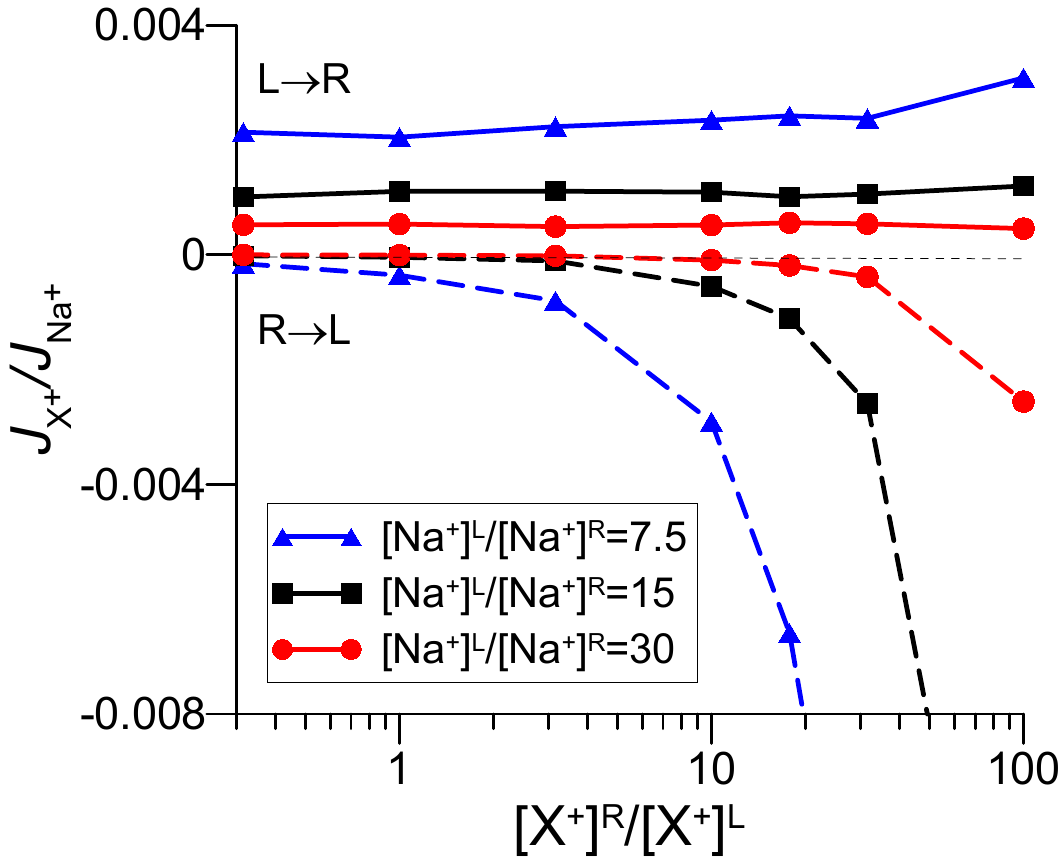}}}
\end{center}
\vfill
\caption{\label{fig5} \small X$^{+}$/Na$^{+}$ flux ratio as a function of the X$^{+}$ concentration ratio for different Na$^{+}$  concentration ratios.
The L-side X$^{+}$ concentration was kept fixed at [X$^{+}$]$^{\mathrm{L}}=1$ mM, so [X$^{+}$]$^{\mathrm{R}}$/[X$^{+}$]$^{\mathrm{L}}$ was changed by changing [X$^{+}$]$^{\mathrm{R}}$.
Panel A shows the net flux, while panel B shows the R$\rightarrow$L (negative, dashed lines) and L$\rightarrow$R (positive, solid lines) components. 
The pore parameters are $Q=-10e$, $R_{\mathrm{p}}=0.24$ nm, and $H=1$ nm.
}
\end{figure*}

\begin{figure}[t]
\begin{center}
\rotatebox{0}{\scalebox{0.75}{\includegraphics*{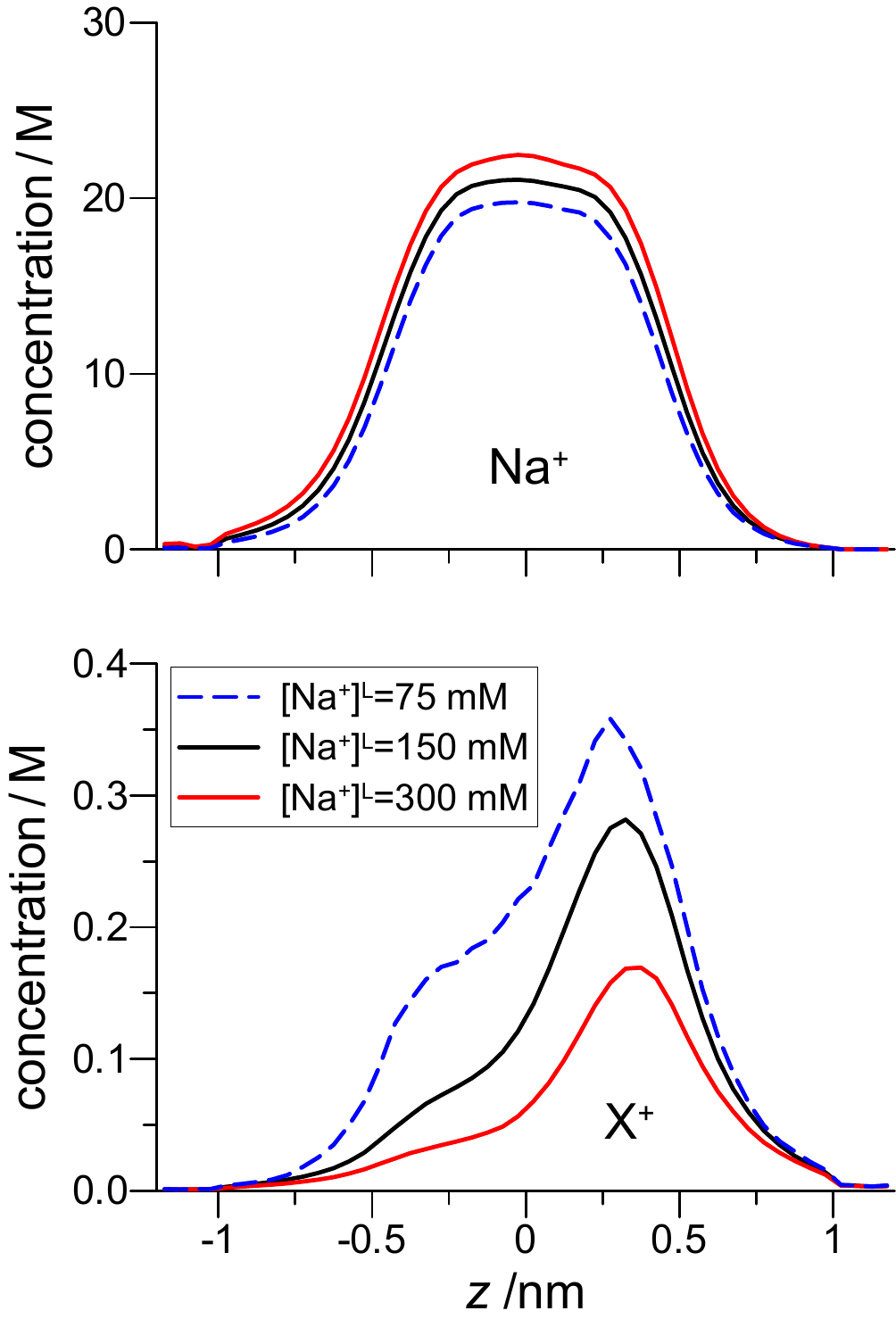}}}
\end{center}
\vfill
\caption{\label{fig6} \small X$^{+}$ and Na$^{+}$ concentration profiles for different L-side Na$^{+}$ concentrations. 
The X$^{+}$ concentrations are $[\mbox{X}^{+}]^{\mathrm{L}}=1$ mM and $[\mbox{X}^{+}]^{\mathrm{R}}=3.16$ mM.
% The Na$^{+}$ concentrations are $[\mathrm{Na}^{+}]^{\mathrm{L}}=150$ mM and $[\mathrm{Na}^{+}]^{\mathrm{R}}=10$ mM; ratio is 15 (L/R). 
The R-side Na$^{+}$ concentration is fixed at 10 mM.
% , while the L-side Na$^{+}$ concentration is varied. 
The pore parameters are $Q=-10e$, $R_{\mathrm{p}}=0.24$ nm, and $H=1$ nm.
}
\end{figure}

Next, we investigated how the co-transport depends on X$^{+}$ and Na$^{+}$ concentration ratios.  
We fixed the Na$^{+}$ concentration on the R side at 10 mM and changed it on the L side in the range 75-300 mM. 
We also fixed the X$^{+}$ concentration on the L side at 1 mM and changed it on the R side in the range 0.316-31.6 mM, the concentration gradient X$^{+}$ ions must fight
 against. 
\ref{fig5}A shows the net X$^{+}$/Na$^{+}$ flux ratio as a function of the X$^{+}$ concentration ratio for different values of $\mbox{[Na$^{+}$]}^{\mathrm{L}}/\mbox{[Na$^{+}$]}^{\mathrm{R}}$. 
\ref{fig5}B shows the L$\rightarrow $R and R$\rightarrow $L components. 
For a given $\mbox{[Na$^{+}$]}^{\mathrm{L}}/\mbox{[Na$^{+}$]}^{\mathrm{R}}$, we observe co-transport if $\mbox{[X$^{+}$]}^{\mathrm{R}}$/ $\mbox{[X$^{+}$]}^{\mathrm{L}}$ is not too large. 
Obviously, if it is too large, the R$\rightarrow $L diffusion dominates over the L$\rightarrow $R co-transport. 
As $\mbox{[X$^{+}$]}^{\mathrm{R}}/\mbox{[X$^{+}$]}^{\mathrm{L}}\rightarrow 1$, the R$\rightarrow $L component (diffusion) goes to zero. 
The L$\rightarrow $R component (co-transport) is constant and does not depend on the X$^{+}$ concentration ratio. 
Consequently, we observe another saturation effect: the X$^{+}$ to Na$^{+}$ flux ratio converges to a limiting value as $\mbox{[X$^{+}$]}^{\mathrm{R}}/\mbox{[X$^{+}$]}^{\mathrm{L}}$ decreases (far left points of \ref{fig5}A). 
This limiting value decreases with increasing Na$^{+}$ concentration ratio.

\begin{figure}[t]
\begin{center}
\rotatebox{0}{\scalebox{0.75}{\includegraphics*{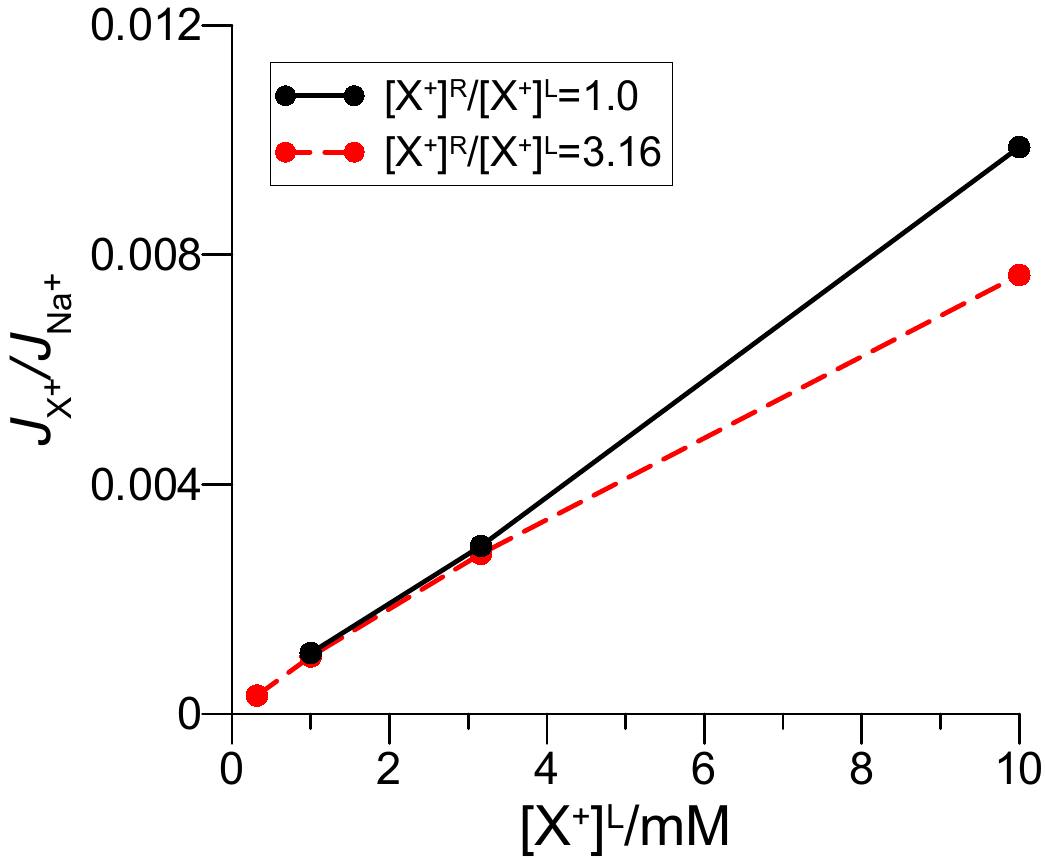}}}
\end{center}
\vfill
\caption{\label{fig7} \small X$^{+}$/Na$^{+}$ flux ratio as a function of the L-side X$^{+}$ concentration for different X$^{+}$ concentration ratios.
The pore parameters are $Q=-10e$, $R_{\mathrm{p}}=0.24$ nm, and $H=1$ nm.
The L- and R-side Na$^{+}$ concentrations are 150 mM and 10 mM, respectively.
}
\end{figure}

To understand why, consider the concentration profiles as [Na$^{+}$]$^{\mathrm{L}}$ increases (\ref{fig6}). 
These show that the concentration of X$^{+}$ decreases substantially with higher [Na$^{+}$]$^{\mathrm{L}}$. 
Apparently, there is a competition between X$^{+}$ and Na$^{+}$ ions for the pore; increasing Na$^{+}$ concentration in the L bath from where it arrives favors adsorbing more Na$^{+}$ and less X$^{+}$ in the pore. 
Also, we varied the L-side X$^{+}$ concentration while keeping the Na$^{+}$ concentrations unchanged. 
Our results (\ref{fig7}) show that the $J_{\mathrm{X}^{+}}/J_{\mathrm{Na}^{+}}$ flux ratio increases approximately linearly with [X$^{+}$]$^{\mathrm{L}}$ because more X$^{+}$ ions are adsorbed in the pore as the L-side X$^{+}$ concentration increases. 
Combined, these results of varying [Na$^{+}$]$^{\mathrm{L}}$ and [X$^{+}$]$^{\mathrm{L}}$ independently indicate that their ratio determines how much X$^{+}$ is in the pore and therefore how much X$^{+}$ is conducted.

Our results have biological implications in that they provide support to the idea that channel-like co-transport through narrow pores can be a kind of transfer mode in transporters.\cite{chen_bj_1993,eisenberg-1996-1,lester_n_1996,sonders_conb_1996,defelice_arp_2007}
The traditional view is that co-transporters behave the same way as those using primary energy (ATPases, for example). 
In this picture, substrates bind to the transporter, induce a conformational change, and this change in configuration results in release of substrates on the other side of the membrane. 
New experimental evidence from fluorescence spectroscopy and electrophysiology, however, shows that current flowing through co-transporters can be orders of magnitudes larger than that possible on the basis of the alternating access model \cite{MolBiolCell}.
These currents are in the range of currents carried by ion channels. 
Moreover, it was reported that the number of the dominant ions far exceeds (10 to a 100 times) the number of co-transported substrate molecules \cite{defelice_n_2004}, in contrast to the fixed stoichiometric mechanism. 
Our simulations are consistent with these properties and show this same range of downhill to uphill ion flux ratio.

Our results also have implications for engineered materials. 
Synthetic nanopores and nanoporous materials have the advantage over biological that their properties (e.g., length $H$, radius $R_{p}$, and charge $Q$) can be more easily manipulated. 
Therefore, our systematic study gives insights into how to optimize these parameters. 
Our simulations also show that only a very small segment of the pore needs to be single-filing. 
This is important for membranes that are usually several microns thick; having sub-nanometer-wide pores spanning the entire membrane would not only be difficult to make, but would also dramatically increase the resistance to ion flow. 
In addition, we showed that large \textquotedblleft uphill\textquotedblright\ ion concentration gradients (more than 10-fold) can still sustain co-transport.
That can be important for applications like amplifying the concentration of a low-concentration analyte molecule. 
One can use a co-transporting membrane to shuttle analyte molecules to higher and higher concentrations to make it easier to analyze them or detect their presence. 
Similarly, one can accumulate an ion concentration gradient for energy storage. 
Or, one can potentially remove contaminating ions (e.g., radioactive ions) with this mechanism.

Nanopores in membranes are becoming small enough that these applications will be possible soon. 
Currently, nanopores can be made to have diameters of $\sim 1$ nm (reviewed by Howorka and Siwy \cite{howorka_csr_2009}), which is almost small enough to force single-filing of ions. 
Moreover, biological ion channels are now being incorporated into these synthetic nanopores\cite{hall_nnt_2010,kocer_bsbe_2012}, including gramicidin A which is a single-file channel.\cite{balme_nl_2011}
Therefore, we suggest that co-transport of ions may be a new possible application of nanopores.

\section{Conclusions}
\label{sec:conclusion}

In conclusion, we have performed DMC simulations for a reduced model of coupled transport in a narrow multiply occupied pore. 
The simplified model made it possible to obtain simulation results with good statistics; the error bars in the figures are about in the size of the symbols. 
At the same time, our model includes the relevant physics to ensure that our results and interpretations are valid.
We found that X$^{+}$ ions can travel uphill using momentum coupling with Na$^{+}$ ions that are present at high concentration (compared to X$^{+}$), driven by their own concentration gradient with normal diffusion. 
Co-transport occurs because thermal motion produces momentum-coupling between X$^{+}$ and Na$^{+}$ ions on the microscopic level.
Macroscopic parameters influence this coupling in various ways.
Geometrical parameters ($R_{\mathrm{p}}$ and $H$) and pore charge ($Q$) determine how strong this coupling is in the crowded bottleneck of the pore.
Na$^{+}$ and X$^{+}$ concentrations both influence the number of X$^{+}$ ions in the pore, and, therefore, the X$^{+}$ flux.
We found that coupling can be established without single filing in a long narrow pore; a short, but narrow opening, where ions are crowded by strong electrostatic attraction exerted by pore charges, is enough.

Voltage was not applied in this study, so the passive diffusion of ions is driven by only the concentration differences between the control cells.
This is the first logical step in studying the phenomenon of co-transport. 
A voltage driving the Na$^{+}$ ions in the direction of their concentration gradient would facilitate co-transport because it would increase the flux of Na$^{+}$. 
Due to momentum transfer, co-transport would be stronger as the drift velocity of Na$^{+}$ ions increases. 
In addition, any voltage favoring Na$^{+}$ conduction would similarly aid X$^{+}$ conduction. 
A voltage of opposite sign, on the other hand, would work against co-transport by decreasing Na$^{+}$ current and moving X$^{+}$ away from the membrane. 
% Applying a voltage, however, would not alter our main conclusions.
Simulations of these effects would be interesting (using one of the electrostatic algorithms of Refs.\ \onlinecite{graf_jpcb_2000,graf_jpcb_2004,cheng_jpcb_2005,hoyles-cpc-115-45-1998,chung-bj-75-793-1998,chung-bj-77-2517-1999,corry-bj-2000,corry-bj-2001,corry-bj-82-1975-2002,chung-bba-1565-267-2002,corry-bj-84-3594-2003,Corry_BBA_1,im_bj_2000,noskov-bj-87-2299-2004,allen_jgp_2004,luo_jpcb_2010,egwolf_jpcb_2010,lee_jcc_2011,lee_bj_2011,kwon_jgp_2011,debiase_jctc_2012,boda-jctc-8-824-2012}) and can be the topic of future studies.
However, the main conclusions of this study would not change. 
Moreover, one of the main goals of our work is new low-energy applications for nanopores and, therefore, we show that co-transport can occur in systems that are completely passive (i.e., where the membrane potential is not fixed and only concentration differences drive the process).

\section*{Acknowledgment}
\label{sec:ack}

We want to thank Lou DeFelice for the conversations that inspired this work.
We are grateful for the valuable discussions with Bob Eisenberg and for drawing our attention to the importance of unidirectional fluxes.
We acknowledge the financial support of the Hungarian State and the European Union under T\'AMOP-4.2.2/B-10/1-2010-0025 and T\'AMOP-4.1.1/C-12/1/KONV-2012-0017.
The support of the Hungarian National Research Fund (OTKA K75132) is acknowledged.

% \appendix
\section{Appendix: Dynamic Monte Carlo}
\label{sec:appendix}

We have chosen the DMC method as introduced by Rutkai et al. \cite{rutkai_jcp_2010} to simulate the movement of X$^{+}$ and Na$^{+}$ ions through our model pore.
The concentrations of these ions are maintained on the two sides of the membrane with Grand Canonical Monte Carlo (GCMC) simulations in control cells.
%  (\textbf{this is the Dual Control Volume (DCV) method \cite{heffelfinger_jcp_1994,im_bj_2000,lisal_jcp_2004}}).
Thus, we simulate steady-state flux.

DMC provides an alternative with many advantages \cite{huitema_jcp_1999,martin_jcp_2001,berthier_pre_2007,rutkai_jcp_2010} over molecular dynamics and Brownian dynamics, for example, shorter computation time, easier handling of hard sphere forces. 
In DMC, the flux of a given ionic species, $J_{i}$, can be computed by counting the particles crossing a predefined reference plane (from L to R and from R to L) in a given MC time interval (time is expressed as the number of trial MC steps in DMC).
In the procedure developed by Rutkai and Krist\'of \cite{rutkai_jcp_2010} to simulate mixtures with DMC, the computed number is divided by the square root of the mass of the given component.
This is the only place, where particles masses enter the calculation.
Choice for the mass of the X$^{+}$ ions would influence the flux ratios, but it does not influence the qualitative trend in the figures.
The X$^{+}$ ions are 7.7 times heavier than Na$^{+}$ ions, in our calculations.
%  (this number corresponds to the serotonin ion).

The basic DMC step is the random particle displacement, in which one particle is chosen from the $N$ particles available in the system with $1/N$ probability, and it is moved into a new position, $\mathbf{r}^{\mathrm{new}}$, with respect to the old position, $\mathbf{r}^{\mathrm{old}}$, with $\mathbf{r}^{\mathrm{new}} = \mathbf{r}^{\mathrm{old}} +  {\bm \xi} r_{\mathrm{max}}$, where $\bm \xi$ is a vector containing three coordinates that are uniformly generated random numbers in the interval $[-1,1]$.
The move is accepted with probability $ p=\min \left\lbrace 1, \exp \left( -\Delta U /kT \right) \right\rbrace $, where $\Delta U$ is the energy-change associated with the movement.
$\Delta U$ may increase, for example, if the electrostatic energy becomes unfavorable or two ions come too close and overlap, resulting in rejection of that new configuration.
The DMC method is based on the assumption that the sequence of configurations generated by the above steps can be considered as a dynamic evolution of the system in time \cite{binder_book}. 
DMC does not generate deterministic trajectories; it reproduces average dynamic properties such as the mean-square displacement. 
Compared to molecular dynamics, DMC does not guarantee an absolute measure of physical time; it only ensures proportionality, which is why it directly provides only relative fluxes.

The choice of the maximum displacement, $r_{\mathrm{max}}$, is a central problem in the DMC method. 
For systems, where every species are modeled explicitly, the value of $r_{\mathrm{max}}$ can be determined from the average free path that a molecule can move toward its neighbors until collision as described in the paper of Rutkai and Krist\'of \cite{rutkai_jcp_2010} in detail.
In this case, the key property determining the value of $r_{\mathrm{max}}$ is the density of the fluid.
The algorithm of Rutkai and Krist\'of \cite{rutkai_jcp_2010} was justified by comparing to results of molecular dynamics simulations.

When we simulate particles moving in an implicit solvent, on the other hand, $r_{\mathrm{max}}$ can be chosen to mimic the stochastic random walk of particles colliding with the solvent molecules.
In this case, the DMC method is more reminiscent of Brownian dynamics simulations.
Consequently, tuning the $r_{\mathrm{max}}$ parameter, DMC simulations can mimic both molecular and Brownian dynamics simulations depending on the presence of implicit degrees of freedom.
Although we observed some (slight) sensitivity of our quantitative results to the value of $r_{\mathrm{max}}$, we did not change its value in our calculations, because we are interested in general qualitative behavior. 
Changing $r_{\mathrm{max}}$ did not influence our qualitative conclusions.

To maintain concentrations on the two sides of the membrane, we apply GCMC in the large, bulk-like containers (called ``control cells'') on the two sides of the membrane \cite{Pohl_mp_1996,im_bj_2000,enciso_mp_2002}.
This is the Dual Control Volumes (DCV) method \cite{heffelfinger_jcp_1994,im_bj_2000,lisal_jcp_2004} to maintain steady-state flux.
Note that the DCV method was applied in the case of ionic systems by Im et al. \cite{im_bj_2000} for the first time.
The control cells are charge neutral on average, although charge can fluctuate in them as individual ions are inserted/deleted in the GCMC steps.

The electrochemical driving force for the passive diffusion of ions is the gradient of the electrochemical potential, $\tilde{\mu}_{i}=\mu _{i}^{0}+kT\ln c_{i}+\mu_{i}^{\mathrm{ex}}+q_{i}\Phi $, where $\mu _{i}^{0}$ is a reference chemical potential, $k$ is Boltzmann's constant, $T$ is temperature, $\Phi $ is the electrical potential, $c_{i}$ is the concentration, $\mu _{i}^{\mathrm{ex}}$ is the excess chemical potential, and $q_{i}$ is the charge of ionic species $i$. 
The electrochemical driving force basically has two components: the gradient of the chemical potential  ($kT\ln c_{i}+\mu_{i}^{\mathrm{ex}}$) and the gradient of the electrical potential ($\Phi$). 

\begin{figure}[t]
\begin{center}
\rotatebox{0}{\scalebox{0.75}{\includegraphics*{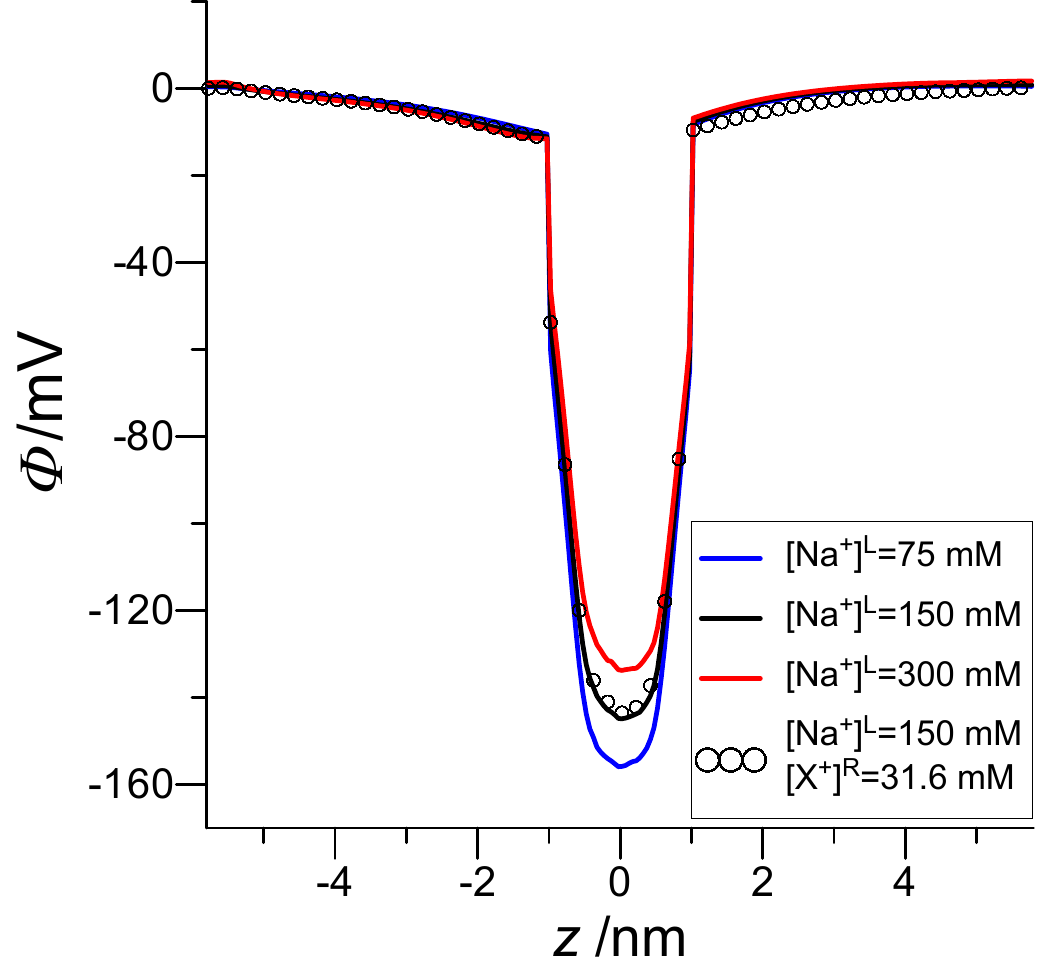}}}
\end{center}
\vfill
\caption{\label{fig8} \small Average electrical potential profiles for the three cases of \ref{fig6} using $[\mathrm{X}^{+}]^{\mathrm{R}}/[\mathrm{X}^{+}]^{\mathrm{L}}=3.16$. An additional curve (symbols) for $[\mathrm{Na}^{+}]^{\mathrm{L}}=150$ mM is shown for a larger X$^{+}$ concentration ratio $[\mathrm{X}^{+}]^{\mathrm{R}}/[\mathrm{X}^{+}]^{\mathrm{L}}=31.6$, where co-transport is not present. 
}
\end{figure}

In this study, we imposed an applied voltage (i.e., electrical potential difference between the baths) of 0 V by enforcing charge neutrality in each of the control cells. 
This is shown in \ref{fig8} with the electrostatic potential profiles for several cases (computed by inserting test charges in the system uniformly). 
The same effect would have been achieved with electrodes at the ends of the baths (but without the charge neutrality condition), something done in many labs with biological ion channels and synthetic nanopores.

Without the explicit electrodes, each control cell approximates a large, well-stirred bath. 
If only a small number of pores are included in the membrane, then this is an excellent approximation because only a small amount of excess charge is moved in that case. 
Moreover, the ion current through these pores will always be small because they are very narrow and thus have high resistance. 
In applications where many pores are in a membrane, this approximation may break down. 
However, the man-made membranes we have in mind are generally micrometers thick and thus have very low capacitance; this would not be true of 3 nm thick biological membranes.
Moreover, flowing fresh electrolyte solution on the side of ion accumulation can be done for some applications. 
Thus, while in principle enough ions can accumulate to create a membrane voltage to stop the co-transport, our large, well-stirred bath approximation is reasonable for many engineering applications.

We used this well-stirred bath approximation because simulating very long pores and large baths is challenging with any simulation technique.
However, using short pores and small baths introduces artifacts of charge accumulation and with it a transmembrane potential. 
While this can happen in biological situations where cell membranes have a large capacitance and the cytoplasm is not at all well-stirred, the applications we have in mind with man-made membranes do not have these properties; baths are generally large so that the ions can quickly diffuse away. 
By using the control cells the way we did, we show the general principle of co-transport (which will occur in longer pores as well), but avoid the artifact of charge accumulation produced by using thin membranes and small baths.

Periodic boundary conditions were used for the control volumes in directions perpendicular to the direction of the transport ($x$ and $y$), while the cell was confined between hard in the $z$ dimension.
GCMC simulations use the chemical potentials as independent variables and apply particle insertion/deletion steps thus simulating a system with fluctuating particle numbers.
This fluctuation, however, occurs around well-defined average values, so the composition (the concentrations of the various species) in the control cells is well-defined.
The chemical potentials of the various species have been determined with the Adaptive GCMC method \cite{malasics-jcp-132-244103-2010}.
The DMC technique coupled to control cells (called DMC+DCV) was used to study transport through ion channels \cite{rutkai-jpcl-1-2179-2010,csanyi-bba-1818-592-2012} and carbon nanopores \cite{Seo_jms_2002}.

The assumption that makes GCMC simulation in the control cells possible is that they are separately in equilibrium.
It is possible, however, to apply GCMC simulations in the transport region too using a local electrochemical potential as introduced in the Local Equilibrium Monte Carlo (LEMC) method \cite{boda-jctc-8-824-2012}.
We also demonstrated that the DMC technique can be coupled to the LEMC method \cite{hato-jcp-137-054109-2012}.

\providecommand*{\mcitethebibliography}{\thebibliography}
\csname @ifundefined\endcsname{endmcitethebibliography}
{\let\endmcitethebibliography\endthebibliography}{}

% \bibliography{transporter,own,dmc-lemc,channel}

\end{document}